\documentclass[10pt,a4paper]{article}
\usepackage{amsmath}
\usepackage{graphicx}
\usepackage{multirow}
\usepackage{epsfig}

\title{MECHANICAL STABILITY PARADIGM FOR PREDICTION OF RPV SERVICE LIFE}

\begin{document}

\maketitle

\begin{center}
S. Kotrechko$^{*1}$, Yu. Meshkov$^1$, I. Nekludov$^2$, V. Revka$^3$, L. Chyrko$^3$\\[.5cm]

$^1$ G.V.Kurdyumov Institute for Metal Physics, NAS of the Ukraine\\
36 Academician Vernadsky Blvd., 03680 Kiev, Ukraine\\[.2cm]
$^2$ National Scientific Center - Institute for Physics and Technology,\\ Kharkiv, Ukraine\\[.2cm]
$^3$ Institute for Nuclear Research, Kiev, Ukraine
\end{center}

\begin{abstract}
In the framework of the engineering version of a Local approach to fracture the criterion of a fracture limit of RPV with crack-like flaw is derived. It has been shown that the level of ductile state stability (mechanical stability) of metal ahead of a crack governs the value of critical fluence. On this basis, the new paradigm of life-time prediction using the condition of exhaustion of mechanical stability of irradiated PV metal has been proposed. The technique of an end-of-life fluence assessment is developed, and predictive capabilities of this approach are demonstrated by the example of WWER-1000 pressure vessels.\\[0.5cm]
\end{abstract}

\noindent*Corresponding author.\\ 
Tel: +380 44 424 13 52\\
Fax: +380 44 424 25 61\\
E-mail:	kotr@imp.kiev.ua (S. Kotrechko)
\section*{Nomenclature}

\begin{tabular}{ll}
$B_h$ 								& is radiation hardening coefficient; \\
$B_{1T}$; $B_{PV}$ 					& crack front length for a 1T compact specimen and a design \\ & crack in the RPV wall; \\
$F_{0i}$ 							& is probability of the crack nuclei instability in {\it i-th} cell; \\
$F_{ni}$, $F_{\Sigma}$ 				& are failure probability of {\it i-th} cell (local probability) and a \\ & whole pre-cracked structure global probability) accordingly; \\
$J_I$ 								& is $J$-integral; \\
$j$ 								& is the parameter of tri-axiality of the stressed state; \\
$K_{Jc}$ 							& is fracture toughness for a CT-1T compact specimen; \\
$K_{Jc}^{PV}$ 						& is fracture toughness for the RPV design crack; \\
$K_{ms}$ 							& is mechanical stability coefficient; \\
$K_I^{PV}$ 							& is stress intensity factor for the RPV design crack at the \\ & critical temperature $T_c^{PV}$; \\
$L$ 								& is relative load; \\
$m$ 								& is radiation hardening exponent; \\
$n$ 								& is work hardening exponent; \\
$P_{ms}$ 							& is mechanical stability parameter; \\
PZ 									& is “process zone”; \\
$q_{\sigma}$ 						& is parameter related to the effect of embrittlement by the \\ & crack ("force equivalent" of the effect of metal embrittlement \\ & by the crack); \\
$T_{cl}$ 							& is critical temperature of the RPV wall fracture; \\
$T_S$ 								& is pressurized thermal shock (PTS) temperature; \\
$\rho_i$ 							& is intensity of nucleating crack formation in an {\it i-th} volume; \\
$\sigma_{0.2}$						& is 0.2\% proof stress for unirradiated RPV metal; \\
$\sigma_{0.2}^{ir}$ 				& is 0.2\% proof stress for irradiated RPV metal; \\
$\sigma_{11i}$ 						& is local tensile stress in {\it i-th} cell; \\
$\sigma_{2}$, $\sigma_{2}^{ir}$ 	& are 2\% offset yield strength for unirradiated and irradiated \\ & metal accordingly; \\
$\sigma_{2c}$ 						& is critical fracture stress for the RPV wall design crack; \\
$\sigma_{f}$ 						& is local fracture stress; \\
$\Phi_c$ 							& is critical fluence for the RPV wall.
\end{tabular} 
\section{Introduction}

Recently, one of the actual problems of fracture research and structural integrity is a use of the physically based approaches instead of empirical engineering ones for a prediction of brittle fracture of the structural components. From the practical point of view this necessity is specified by increasing the technical and economical factors of a component operation thanks to decreasing a conservatism of strength calculation. This problem arises especially for the reactor pressure vessels as, on the one hand in this case the RPV metal embrittlement is most brightly appeared for the operation, and on the other hand the RPV life time extension due to decreasing a conservatism of calculations gives a considerable economical profit. However, the acceptable decreasing a conservatism of calculations is possible only by means of an use of reliable approaches to the prediction of a metal brittleness level which is depend on both a metal microstructure and loading conditions. Such methods, on the one hand reflect the real brittle crack initiation in an irradiated material considering the tri-axial local stresses in the vicinity of crack-like flaws in the RPV, and on the other hand give a criteria of the limit state in the form suitable for engineering calculations.

A conventional approach to the prediction of RPV service life consists in using a temperature margin between a critical brittleness temperature $T_K$ for the specified period of RPV operation and its maximum allowed value $T_K^a$. Previous temperature $T_K$ is estimated using surveillance test data and the latest one $T_K^a$ is defined on the base of strength calculations considering a presence of sharp cracks in the RPV and pressurized thermal shock (PTS) conditions. The RPV metal radiation embrittlement rate is evaluated by a shift of the critical brittleness temperature, $\Delta T_F$, obtained from the surveillance tests. As usual the temperature $T_K$ and the radiation shift $\Delta T_F$ are determined using the Charpy impact (dynamic) test results [1, 2]. Unfortunately such kind of tests does not reflect the real loading conditions which take a place in the RPV metal. In addition Charpy impact test data give too conservative values of the critical brittleness temperature, i.e. highly underestimate the RPV radiation life time. Nowadays in this relation there is tendency to use a Master curve approach for a prediction of the RPV service life. This approach is used to perform static fracture toughness tests using the pre-cracked Charpy specimens and evaluate the reference temperature $T_0$ which then used for a determination of the initial critical brittleness temperature. However the radiation shift $\Delta T_F$ is determined again using Charpy impact test data. So, an application of fracture mechanics methods have only led to a change in evaluation of the initial critical brittleness temperature, but a \textit{paradigm} of the RPV life prediction itself has remained invariable. As before, the value of critical brittleness temperature is used as a criterion of the metal ability to resist the brittle fracture at RPV loading.

In the paper [9] it is shown a possibility to estimate the radiation embrittlement rate for RPV steels not using an indirect parameter as a shift $\Delta T_F$, but by means of a direct method of the exhaust of irradiated metal plasticity in a local area ahead of the crack tip. In the papers [9-10] an analysis of physical phenomena of cleavage fracture initiation in the vicinity of a macro crack tip has been performed for metals and the new parameters of the material ability to resist a ductile to brittle transition have been obtained. A new approach to the prediction of the fracture limit for structural components with crack-like flaws was suggested considering the results of this analysis.

The aim of this work is to present a new paradigm for the prediction of the fracture limit for a reactor pressure vessel containing crack-like flaws and the estimation of RPV service life.
\section{Theoretical basic}

As is well known in the most cases a brittle fracture of a structural component with a crack is initiated within the local region in front of the crack (notch). So the most adequate description of this process is possible within the framework of Local approach (LA) to fracture [12-15]. The main advantage of this approach is possibility to consider both specific features of the micro mechanism of fracture initiation and effects related to the high strain gradients and tri-axial loading within a local area ahead of the crack tip [15].

The cleavage fracture initiation inside of the local region, “process zone” (PZ), in front of the macroscopic crack is a stochastic process. It consists in formation and instability of the crack nuclei in the PZ. A generalized statistical criterion of the fracture initiation of a pre-cracked solid is the following [14]:

\begin{equation}
	F_{ni}(\sigma_{11i}) = 1 - [1-F_{0i}(\sigma_{11i})]^{\rho_i V_i}
\end{equation}

\begin{equation}
	F_{\Sigma} = 1 - \prod_{i=1}^{i=M} [1-F_{ni}(\sigma_{11i})]
\end{equation}
where $F_{0i}$ is the probability for instability of the crack nuclei in {\it i-th} cell~\footnote[1]{Finite element method is usually employed to calculate stresses and strains ahead of a crack, so, "cell" means "finite element"} at the value of local stresses $\sigma_{11i}$ acting in this cell; $V_i$ is the cell volume; $\rho_i$ is the rate of the crack nuclei generation within the unit volume at the specified value of equivalent local plastic strain $\bar{e_i}$ in {\it i-th} cell $\rho_i = (dN/d\bar{e_i}) \cdot (1/V_i)$~\footnote[2]{The value of the CN generation rate within the volume unit is used instead of the cumulated density ($N_i/V_i$) because cleavage fracture of metals may be initiated only by the cracks, which become unstable at the moment of their nucleation [15, 16].}, where $N_i$ is the general number of the crack nuclei formed in the $\bar{e_i}$  deformed cell); $F_{ni}$ and $F_{\Sigma}$ is the probability for fracture of {\it i-th} cell (local probability) and a whole structure (global probability) accordingly; $M$ is the number of cells in the PZ volume ahead of the crack (notch).

A solution of these combined equations at a specified confidence level of the global probability for fracture, $F_{\Sigma}$, enables to estimate the critical value of a local tensile stress, $\sigma_f$, at the point in front of the crack where a maximum probability, $F_{ni}^{max}$, for cleavage initiation is reached (Fig. 1). This magnitude of $\sigma_f$ is obtained for the pre-cracked solid at the critical value of $K_I$, at which the failure is occurred with a specified global probability $F_{\Sigma}$.

In the general case, to find the value of a fracture load, two problems must be solved, namely: (i) to obtain a distribution of local stresses and strains in a vicinity of the crack tip for the specified $K_I$ and (ii) to estimate the local $F_{ni}$ and global $F_{\Sigma}$ probability for failure initiation. For solving this problem in a conventional version of the Local approach the Weibull distribution is used instead of the statistical criterion (1) that enables to get the relation between $F_{ni}$ and $\sigma_{11i}$ in the explicit form [13, 14]. In multi-scale version of local approach proposed in [15-17] the probability for fracture in {\it і–th} cell, $F_{ni}$, and the rate of crack nuclei generation, $\rho_i$, are determined by means of the analysis of micro processes of the formation and instability of the crack nuclei in a polycrystalline aggregate.

However, both conventional and the proposed versions of a Local approach is rather sophisticated for engineering applications. So, a simplified version of the local approach is suggested in [9]. A specific feature of this version is that it enables, on the one hand, to consider for peculiarities of a micro-mechanism of cleavage fracture initiation in the local region at the macro crack tip and, on the other hand, to express the criterion of cleavage niitiation in terms of global parameters which are evaluated by means of the calibration procedures. In this version of Local approach the generalized stochastic criterion (1) and (2) is expressed in stresses that is more convenient for the engineering calculations:

\begin{equation}
	P_{ms} = \frac{\sigma_f}{\sigma_{11}} = 1
\end{equation}
where $\sigma_f$ is the most probable value of the local stress of fracture initiation (Fig. 1); $\sigma_{11}$ is local tensile stresses acting within a PZ where the probability of cleavage fracture initiation is maximum.

This criterion differs from the similar criteria of brittle fracture by $\sigma_f$ stress properties. The $\sigma_f$ value is not the material’s constant. It depends not only on structure but also on the confidence level of a global probability for failure initiation $F_{\Sigma}$. In addition the $\sigma_f$ stress depends on the PZ volume at the specified value of $F_{\Sigma}$. As a result, the parameter $\sigma_f$ changes depending on the crack front length and loading value which govern PZ size.

The parameter $P_{ms}$ enables to consider the feature of cleavage fracture initiation which is that a plastic strain is a necessary condition for fracture. The crack nuclei are generated only during plastic flow, and their instability results in the cleavage fracture initiation. The latest occurs when the tensile stresses reach a critical value. So, a local plastic strain can realize without a cleavage failure (stable plastic state, $P_{ms} > 1$) or leads to the instability of crack nuclei at the time of their formation (unstable plastic state, $P_{ms}~\leq~1$) depending on loading conditions. Therefore a degree of excess of the parameter $P_{ms}$ over a unity specifies a stability level of the metal plastic state in PZ at the given loading conditions (load, temperature, fluence etc.) [9-11]. Higher the value of $P_{ms}$ in comparison to a unity, higher the level of stability of plastic state, and, respectively, higher the allowable radiation hardening value for the RPV metal.

In criterion (3) the local characteristics $\sigma_f$ and $\sigma_{11}$ are used. A transition from local to global parameters was realized in [9-11]. Therefore the criterion of fracture initiation may be presented as follows:

\begin{equation}
	P_{ms} = \frac{K_{ms}}{q_{\sigma}} = 1
\end{equation}
where $K_{ms}$ is the coefficient of stability of the ductile state at uniaxial tension (shortly – "mechanical stability coefficient"):

\begin{equation}
	K_{ms} = \frac{R_{MC}}{\sigma_Y \cdot (e_c/e_Y)^n}
\end{equation}
$R_{MC}$ is the brittle strength of metal which is experimentally determined as a minimum value of brittle fracture stress over the "ductile to brittle transition" temperature range under uniaxial tension [9, 18]; $\sigma_Y$ and $n$ are the yield strength and the work hardening exponent, respectively; $e_Y$ is the plastic strain correspond to the yield strength (for the structural steels $e_Y \approx 0.2\%$ and $\sigma_Y = \sigma_{0.2}$); $e_c$ is the value of plastic strain at the point where the probability of cleavage failure initiation is maximum (i.e. at stress $\sigma_f$); $q_{\sigma}$ is the parameter characterising the value of embrittlement of metal within the local region ahead of a crack ("force equivalent" of the effect of metal embrittlement by the crack). By the definition:

\begin{equation}
	q_{\sigma} = \frac{j}{k_v}
\end{equation}
where $j$ is the parameter of tri-axiality of stressed state in front of a crack at the point where $\sigma_f$ is reached:

\begin{equation}
	j = \frac{\sqrt{2} \cdot \sigma_{11}}
		{\sqrt{(\sigma_{11}-\sigma_{22})^2 + (\sigma_{22}-\sigma_{33})^2 + (\sigma_{11}-\sigma_{33})^2}}
\end{equation}
Coefficient $k_v$ is the measure of the "local" scale effect. It characterises the degree of excess of the local stress $\sigma_f$ over the value of brittle strength $R_{MC}$ for a standard tensile specimen having the volume of $V= 1000 \ \textrm{mm}^3$:

\begin{equation}
	k_v = \frac{\sigma_f}{R_{MC}}
\end{equation}
The parameter $q_{\sigma}$ indicates how much times the level of stability of a plastic state, $P_{ms}$, in front of the crack in PZ less than the similar value, $K_{ms}$, at {\it uniaxial} tension.

As it was reported in [9] for the RPV steels $e_c \approx 0.02$~\footnote[3]{It is needed to emphasize that at $n \leq 0.1$ that is typical for the structural steels the accuracy in a determination of $e_c$ affects $K_{ms}$ insignificantly}. In this case:

\begin{equation}
	K_{ms} = \frac{R_{MC}}{\sigma_{0.2} \cdot 10^n}
\end{equation}.
It is known that for high grade RPV steels like 2Cr-Ni-Mo-V and А533 the neutron fluence does not influence on a brittle strength (i.e. $R_{MC}$) within a fluence range up to $\sim 10^{24} \ \textrm{m}^2$. It means the main cause of embrittlement in this case is a radiation hardening. According to (9) the decrease of the mechanical stability coefficient, $K_{ms}$, caused by neutron irradiation is related to the increase of a yield stress of irradiated metal. As usual a radiation induced increase in the yield stress is described by the power law:

\begin{equation}
	\sigma_{0.2}^{ir} = \sigma_{0.2} + B_h \cdot \bigg(\frac{\Phi}{10^{22}}\bigg)^m
\end{equation}
where $\sigma_{0.2}^{ir}$ is the yield stress of irradiated metal; $B_h$ is the radiation hardening coefficient; $m$ is then exponent. For the RPV steels $m \approx 0.33 - 0.51$ [19-21].
Substituting (5) to (4) and accounting for (10), and supposing that $e_c = 0.02$ and $e_Y = 0.002$ one can obtain a criterion of fracture at the specified fluence, $\Phi_c$, for the RPV wall with a crack-like flaw:

\begin{equation}
	\frac{R_{MC}}
	{q_{\sigma} \Bigg[\sigma_{0.2} + B_h \cdot \bigg(\frac{\Phi}{10^{22}}\bigg)^m \Bigg] \cdot 10^{n(\Phi_c)}} = 1
\end{equation}
Equation (11) determines the key factors which specify the RPV radiation life time, however, for the engineering application it is reasonable to simplify this criterion by eliminating $R_{MC}$. As it was shown in [9, 10] for the specimen with a crack the value $q_{\sigma}$ numerically equal to $K_{ms}$ at the critical temperature, $T_c$, when fracture occurs at the specified stress $K_I$. Really, according to (4) the fracture initiation condition is the following

\begin{equation}
	K_{ms}(T_{cl}) = q_{\sigma}
\end{equation}
Substituting (12) to (11) with account of the invariance of $R_{MC}$ to irradiation dose and equation (9) gives:

\begin{equation}
	\frac{\sigma_{2c}} 
	{\Bigg[\sigma_{0.2} + B_h \cdot \bigg(\frac{\Phi}{10^{22}}\bigg) \cdot 10^{n(\Phi_c)} \Bigg]} = 1
\end{equation}
where $\sigma_{2c}$ is the critical value of  fracture stress for the un-irradiated specimen at 2\% local strain and the temperature $T_{cl}$ (numerically defined as $\sigma_{2c} = \sigma_{0.2}(T_{cl}) \cdot 10^{n(T_{cl})}$).
Physical meaning of the critical stress $\sigma_{2c}$ is that its value is equal to a local fracture stress $\sigma_f$ with accuracy within a parameter $j$. Really, substituting (8) to (6) and accounting for (9) and (12), gives:

\begin{equation}
	\sigma_{2c} = \frac{\sigma_f}{j}
\end{equation}
It should be noticed that denominator in the equation (13) is equal to $\sigma_{2}^{ir}$ at the desired value of fluence $\Phi_c$. It enables to present a criterion (13) as:

\begin{equation}
	\sigma_{2}^{ir}(\Phi_c) = \sigma_{2c}
\end{equation}
According to (15) to find a critical value of the fluence $\Phi_c$ at the specified load, $K_I$, and the temperature (for example, at PTS temperature $T_S$), it is required to have a dependence of 2\% offset yield strength for irradiated metal, $\sigma_{2}^{ir}$, on the fluence and to know the critical level of $\sigma_{2c}$ for RPV at the specified value of $K_I$.
Equation (13) enables to get implicitly an expression for the critical fluence $\Phi_c$:

\begin{equation}
	\Phi_c = \left[\frac{\sigma_{2c} \cdot 10^{n-n(\Phi_c)} - \sigma_{2}}{B_h \cdot 10^{n}} \right]^{\frac{1}{m}} \cdot 10^{22}
\end{equation}
where $\sigma_{0.2}$ is the yield stress at 2\% strain and the PTS temperature $T_S$ ($\sigma_2 = \sigma_{0.2}(T_S)\cdot10^{n(T_S)}$ ); $n$ and $n(\Phi_c)$ are the work hardening coefficients at the PTS temperature for un-irradiated and irradiated steel, accordingly. The coefficient $n$ depends on $\Phi_c$ so equation (16) is nonlinear in regard to $\Phi_c$. However, an explicit form of the approximate expression for $\Phi_c$ may be obtained accounting for the empirical fact that a deformation hardening coefficient $n$ at the PTS temperature $T_S$ for irradiated metal and at the critical temperature $T_{cl}$ for un-irradiated material is approximately the same:

\begin{equation}
	n(\Phi_c) \approx n(T_{cl})
\end{equation}
where $n(T_{cl})$ is the deformation hardening coefficient for un-irradiated material at the critical temperature $n(T_{cl})$.
Correctness of such approximation results from a relation between the value of $n$ and the steel strength [9]. In this approximation:

\begin{equation}
	\Phi_c = \left[\frac{\sigma_{2c} \cdot 10^{n-n(T_{cl})} - \sigma_{2}}{B_h \cdot 10^{n}} \right]^{\frac{1}{m}} \cdot 10^{22}
\end{equation}
Usefulness of this equation consists first of all in the fact that it enables in an explicit form to show which mechanical parameters determine the RPV radiation life time and to what extent they affect it. According to (18), not only susceptibility of the RPV steel to radiation embrittlement (parameters $B_h$ and $m$) effects significantly the fluence $\Phi_c$, but also the strength of unirradiated metal, $\sigma_{2}$, and it ability to resist the cleavage fracture initiation in a local area at the crack tip (parameter $\sigma_{2c}$). In the relation to an existing tendency to the increasing the RPV steel strength it should be noted that increasing of $\sigma_{2}$ does not results in the decreasing of $\Phi_c$ only if this will be accompanied with a relevant increase of the brittle strength, $\sigma_{2c}$. A detailed consideration of the microstructure factors influencing on $\sigma_{2c}$ and the technological ways to increase this parameter is stated in [9]. Moreover, it is needed to emphasize that a stress $\sigma_{2c}$ depends on the crack front length and the value of $K_I$ under a pressurized thermal shock.
\section{Prediction of the critical fluence $\Phi_c$}

A key point in the method of the $\Phi_c$ fluence determination is to find out the critical strength $\sigma_{2c}$. Idea of this method is shown in fig.2. First, the pre-cracked Charpy specimens are tested to get Master curve for CT-1T compact specimens with $B_{1T} = 25.4$ mm thickness. Then the $K_{Jc}^{PV}$ values are calculated for a design RPV crack with a length of $B_{PV} = 150$ mm. 

\begin{equation}
	K_{Jc}^{PV} = 1.096 \cdot (K_{Jc}^{1T} - K_{min}) \cdot 
	\left[-\frac{B_{1T}}{B_{PV}} \ln(1-F_{\Sigma}) \right]^{1/4}
\end{equation}
where $F_{\Sigma}$ is the confidence level for fracture probability (as usual a value of 0,05 is used for the calculation), $K_{min} = 20 \ \textrm{MPa}\sqrt{\textrm{m}}$.

The $\sigma_{0.2}$ and $n$ parameters are defined using the tension test data for the round surveillance specimens to get a temperature dependence of the strength $\sigma_{2}$. The critical strength $\sigma_{2c}$ is estimated as a value of $\sigma_{2}$ at a critical temperature $T_{cl}$ (fig.2). The $T_{cl}$ value depends on a stress intensity factor $K_I$ which the critical fluence $\Phi_c$ is calculated for. Therefore the temperature $T_{cl}$ is found out as an intercept of the temperature dependence of fracture toughness $K_{Jc}$ and the temperature dependence of $K_I$ [9]:

\begin{equation}
	K_I = \sqrt{\frac{E \sigma_{0.2}}{1-\nu^2} \ bL}
\end{equation}
where $E$ is the Young modulus, $\nu$ is the Poisson ratio, $L$ is the dimensionless parameter of load:

\begin{equation}
	L = \frac{1}{M} = \frac{J_I}{b \sigma_{0.2}}
\end{equation}
where $M$ is dimensionless parameter from ASTM E1921, $b$ is specimen ligament, $J_I$ is the value of $J_I$ - integral.
\newline An absolute value of $L$ or $J_I/\sigma_{0.2}$  is selected in such a way that a $K_I$ value lies within $65-100 \ \textrm{MPа}\sqrt{\textrm{m}}$ at the pressurized thermal shock temperature $T_S$. The dependence of $\sigma_{2}^{ir}$ on the fluence (fig.3) is plotted to ascertain the critical fluence $\Phi_c$ for the temperature $T_S$.

\begin{equation}
	\sigma_{2}^{ir} = \sigma_{0.2}(\Phi) \cdot 10^{n(\Phi)}
\end{equation}
The dependence of $\sigma_{0.2}$  on the fluence is calculated by formulae (10). An empirical relationship obtained in [10] can be used to plot $n$:

\begin{equation}
	n = \frac{\alpha}{\sigma_{0.2}^{\beta}}
\end{equation}
where $\alpha =3.87$ and $\beta = 0.65$ for the 2Cr-Ni-Mo-V type RPV metal. According to equation (15) the value $\Phi_c$ is defined as a fluence which strength $\sigma_{2}$ reaches the critical value $\sigma_{2c}$ at (fig.3).

Finally it is needed to note that necessity to use the Master curve methodology is caused by a small number of surveillance specimens and the use of small size specimens (a thickness $B = 10$ mm) to estimate the $K_{Jc}$ values for a crack front length of 150 mm. Generally it should be emphasized that an application of the Master curve method for the test data analysis needed for calibration procedures in the Local approach is perspective for an engineering application. In particular such efforts were being undertaken in the work [22].

The surveillance specimens for the 2Cr-Ni-Mo-V type RPV steel and welds are used as object under study (table 1). The specimens were irradiated in the WWER-1000 power reactors. Two different types of specimens were tested: (i) pre-cracked Charpy and (ii) standard tensile specimens. In the first case the temperature dependencies of $K_{Jc}^{1T}$ were plotted according to the Master curve method (ASTM E1921 standard).

The static tension tests were performed at the room temperature and $+350^{\circ}$C. The temperature dependences of yield strength $\sigma_{0.2}$ for the material before and after irradiation were been fitted by the conventional relationship:

\begin{equation}
	\sigma_{0.2} = \sigma_{0.2}^{*} + C_1 \exp[-C_2(T + 273) + C_3 \ln(\dot{e})] - 49.6
\end{equation}
where $\sigma_{0.2}^{*}$ is yield strength at room temperature, $C_1 = 1033$ MPa, $C_2 = 0.00698 \cdot T^{-1}$, $C_3 = 0.000415 \cdot T^{-1}$, $\dot{e} = 0.004 \ c^{-1}$.
\newline The value of work hardening exponent, $n$, for examined temperature range was determined according to (23). On the base of this data the temperature dependences of stress $\sigma_{0.2}$ are plotted for both un-irradiated and irradiated conditions. Radiation embrittlement data is fitted by the expression (10) with a power factor $m = 1/3$. The values of the radiation embrittlement coefficient $B_h$ are shown in table 2. The temperature $T_{cl}$ is determine for the relative load value $L = 0.00084$ ($M = 1190.5$; $J_I/\sigma_{0.2} = 0.0365$ mm) for the reactor pressure vessel with a design crack length. For this $L$ the $K_I$ value for a design crack length of 25.4 mm in the RPV wall with a thickness of 200 mm is 66 MPa$\sqrt{\textrm{m}}$ at the PTS temperature $T_S = 56^{\circ}$C~\footnote[4]{For the calculations semi-elliptical crack have been approximated by a through crack. It gives a conservative uncertainty.}. 
\section{Results and discussion}

In fig. 3 the dependences of yield strength $\sigma_{2}$ on neutron fluence are given for two WWER-1000 RPV welds (W1 and W2). The critical stress $\sigma_{2c}$ is plotted on the same figure for the RPV metal both in the initial condition and after irradiation to the difference fluences. That is interesting the critical stress $\sigma_{2c}$ depends on neutron fluence $\Phi$. Meanwhile a tendency of $\sigma_{2c}$ stress increasing with neutron fluence increasing is observed (table 2). An analysis of this effect is a subject for a separate publication. It should be noted that the critical fluence $\Phi_{c}$ calculated by an approximate relation (18) is in good agreement with a numerical solution of the nonlinear equation (16). Generally the data in Table 2 confirm that RPV service life is limited by the maximum allowable fluence $\Phi_{c}$ for welds.

It is needed to emphasize that a critical fluence $\Phi_{c}$ depends also on the confidence level for fracture probability at a relative load $L = J_I/(\sigma_{0.2}\cdot b)$ and a crack size. This is caused by the statistical nature of the cleavage fracture initiation in PZ that results in a dependence of local fracture stress $\sigma_{f}$ , and it means $\sigma_{2c}$, on the confidence level for fracture probability. According to Table 3 this effect is needed to be taken into consideration in the calculation since it may lead to the considerable (near two times) change of RPV service life estimation.

In the framework of suggested approach it is simply enough to take into account the influence both of the load and crack size on the critical value of $\Phi_{c}$. According to equations (21) и (20) the increasing a crack length (the decreasing a ligament $b$ in equation (21)) or the increasing the load at the PTS event leads to an increase of $K_I$ that causes a growth of temperature $T_{cl}$ (fig. 2 and decrease in the value of acceptable irradiation dose, respectively ). For example for weld W1 an increase of the relative load from $L = 2.1\cdot 10^{-4}$ ($J/\sigma_{0.2} = 0.0365$ mm, $K_I = 73$ MPa $\sqrt{\textrm{m}}$) to $L = 3.0 \cdot 10^{-4}$ ($J_I/\sigma_{0.2} = 0.052$ mm, $K_I = 87$ MPa $\sqrt{\textrm{m}}$) results in decreasing the critical fluence from $\Phi^{5\%} = 110 \ \textrm{m}^{-2}$ to $\Phi^{5\%} = 70 \ \textrm{m}^{-2}$.

The crack front length influences on $\sigma_{2c}$ due to a local size effect. In order to quantitative this effect it is necessary to build by Eq.19 the temperature dependence of $K_{Jc}$ for the specified value of B and estimate   for its value (fig. 2). In fig. 4a dependence of $\Phi_{c}^{5\%}$ on the crack front length at fixed level of $J_I/\sigma_{0.2} = 0.0365$ mm is shown.

Therefore, from analysis of initiation of cleavage fracture of RPV metal within the local region ahead of a crack-like flaw tip, it follows that the value of critical fluence $\Phi_{c}^{5\%}$ is specified by the condition of assessment by strength of metal $\sigma_{2}$  it critical level $\sigma_{2c}$. Using of these two characteristics and parameters of irradiation hardening of metal gives us possibility to describe the effect of three main factors (mechanical properties of RPV metal, loading level and crack-like flaw sizes) on the value of RPV life time within the framework of unique approach.

\section{Conclusion:}
\begin{enumerate}
	\item The RPV end-of-life fluence can be estimated not by an indirect parameter which is a shift of the critical brittleness temperature of a surveillance specimen, but by the condition of exhaustion of mechanical stability of an irradiated PV metal in the local area at the vicinity of the crack tip in the RPV wall.
	\item In the framework of a suggested methodology the critical fluence for the reactor pressure vessel is determined by a comparison of 2\% offset proof stress for the irradiated metal with its critical value of $\sigma_{2c}$, at which the cleavage fracture initiation ahead of the crack tip in the RPV wall occurs.
	\item A dependence of the critical fluence on a crack front length and a load level at the pressurized thermal shock is determined by an impact of these factors on the critical stress $\sigma_{2c}$.
\end{enumerate}

\section*{Acknowledgements}
This work is executed due to financial support of "State Integral Program of Fundamental and Applied Researches of the Problems of Use of Nuclear Materials and Radiation Technologies for Economic Development", Project K-3-8.


\newpage
\begin{figure}[h]
	\includegraphics{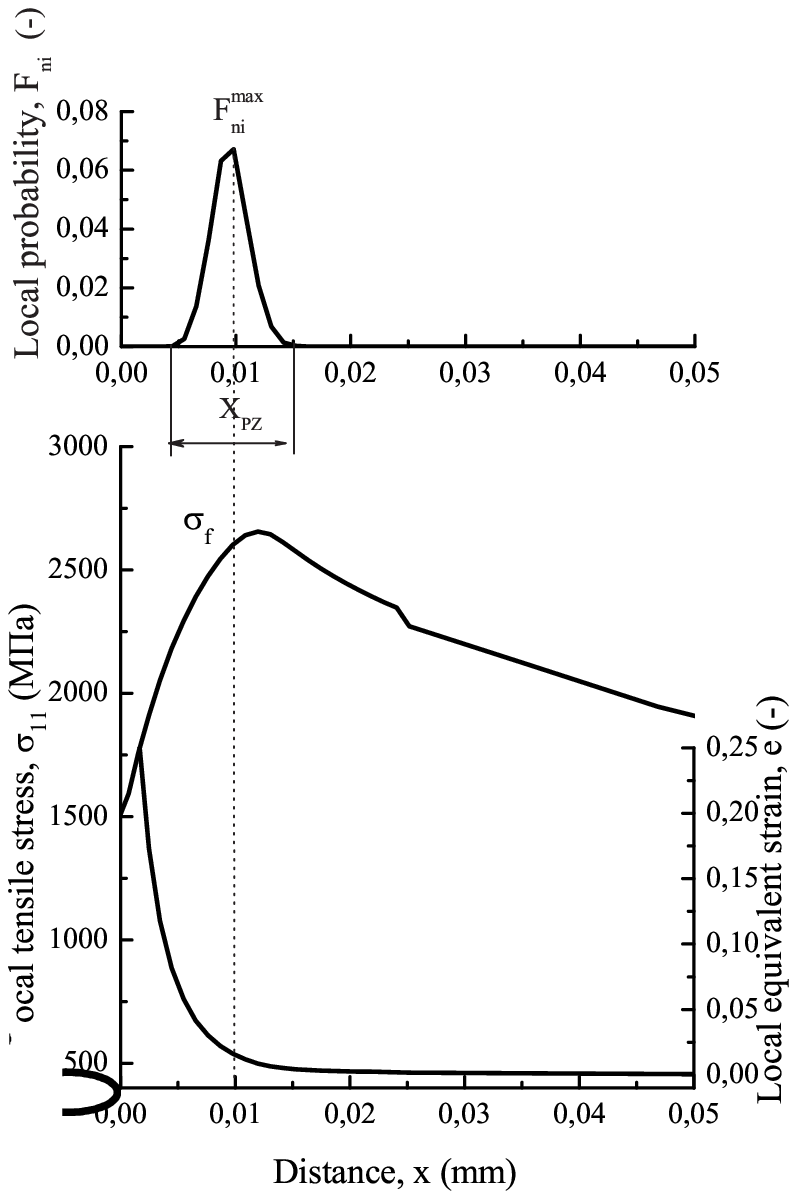}
	\caption{Distribution of local stress, $\sigma_{11}$, equivalent local plastic strain, $\bar{e}$, and $F_{ni}$ is the local probability of fracture initiation ahead of a crack tip in pre-cracked Charpy surveillance specimen at such condition: temperature is equal $-140^{\circ}$C, $K_{Jc} = 30 \ \textrm{MPa}\sqrt{\textrm{m}}$ and probability of overall fracture $F_{\Sigma} = 0.63$; $X_{PZ}$ is size of “process zone” in minimum cross-section of specimen.  \label{Fig:1}}
\end{figure}

\newpage
\begin{figure}[h]
	\centering{
	\includegraphics{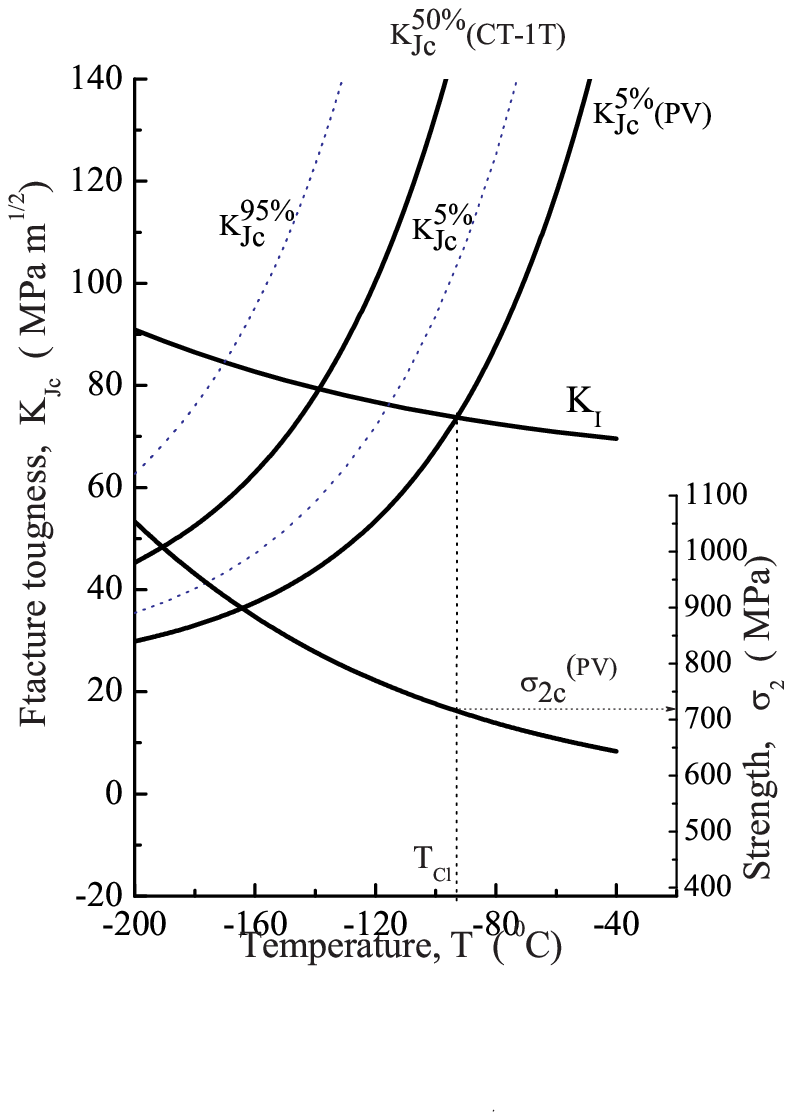}
	\caption{Temperature dependences of fracture toughness $K_{Jc}$ of standard specimen CT-1T at fracture probability  tolerances 5\%, 50\% and 95\%; $K_{Jc}^{5\%}$(PV) is the fracture toughness for the RPV design crack at probability 5\%; $\sigma_{2c}$ is the value of critical stress for RPV wall with crack; $K_{I}$ is loading value; $T_{cl}$ is critical temperature for RPV with crack (weld metal - W1). \label{Fig:2}}}
\end{figure}

\newpage
\begin{figure}[h]
	\centering{
	\includegraphics{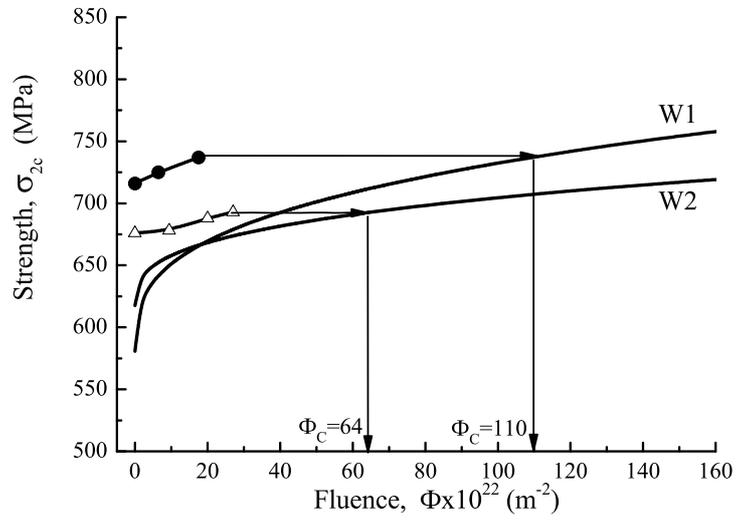}
	\caption{Strength dependence of the weld metals W1 and W2 on neutron irradiation dose at $T_S = +59^{\circ}$C; $\sigma_{2c}$ is the critical stress  for weld metal -- $\bullet$ for W1 and $\Delta$ for W2 –- for the RPV design crack; $J_I/\sigma_{0.2}=0.0365$ mm and fracture probability that equals 5\%. \label{Fig:3}}}
\end{figure}

\newpage
\begin{figure}[h]
	\centering{
	\includegraphics{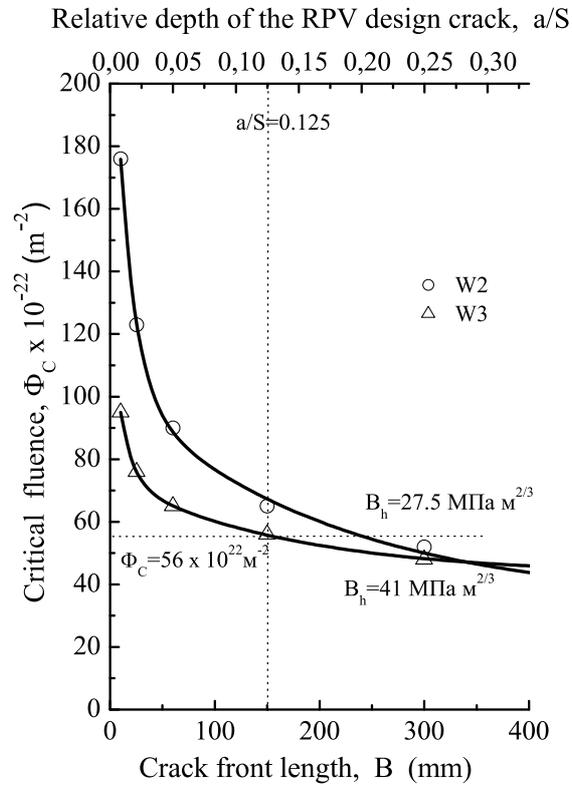}
	\caption{Dependence of the value of critical fluence $\Phi_c$ on the relative depth $a/S$ ($S = 200$ mm is reactor wall depth) and crack front length $B$.  \label{Fig:4}}}
\end{figure}

\newpage
\renewcommand{\arraystretch}{1.5}
\begin{table}[pH] 
\caption{{\bf Chemical composition of analysed PVR metal (weight \%)} Base metal -- steel 15X2NMFA of different fusion (BM) and weld metal (WM)} 
\begin{center} 
\begin{tabular}{|l|c|c|c|c|c|c|c|c|c|c|} \hline
PVR metal	& C 	& Si 	& Mn 	& Cr 	& Ni	& Mo 	& Cu	& S 	& P 	& V		\\ \hline
BM 1		& 0.17	& 0.23	& 0.46	& 2.05	& 1.26	& 0.52	& 0.10	& 0.012	& 0.010	& 0.09	\\ \hline
WM 1		& 0.07	& 0.37	& 0.74	& 1.86	& 1.67	& 0.68	& 0.04	& 0.008	& 0.006	& ---	\\ \hline
BM 2		& 0.17	& 0.21	& 0.44	& 1.99	& 1.19	& 0.56	& 0.12	& 0.009	& 0.010	& 0.09	\\ \hline
WM 2		& 0.09	& 0.26	& 0.93	& 1.7	& 1.74	& 0.56	& 0.05	& 0.010	& 0.007	& ---	\\ \hline
WM 3		& 0.065	& 0.26	& 0.91	& 1.83	& 1.82	& 0.61	& 0.03	& 0.009	& 0.008	& 0.01	\\ \hline
BM 4		& 0.15	& 0.28	& 0.46	& 2.04	& 1.14	& 0.55	& 0.05	& 0.011	& 0.009	& 0.10	\\ \hline
WM 5		& 0.06	& 0.40	& 0.94	& 1.70	& 1.70	& 0.67	& 0.04	& 0.012	& 0.007	& ---	\\ \hline
\end{tabular} 
\end{center} 
\end{table}

\newpage
\renewcommand{\arraystretch}{1.5}
\setlength{\tabcolsep}{4pt}
\begin{table}[pH] 
\caption{Mechanical properties of RPV material and characteristics of fracture of RPV wall with tolerance of $P_f=50\%$ at $J/\sigma_{0.2}=0.0365$} 
\vspace{0.5cm}\hspace{-3.5cm}
\begin{tabular}{|l|c|c|c|c|c|c|c|c|c|c|c|c|c|c|c|}
\hline
\multirow{3}{*}{Mater.}
& $R_{MC}$,	& $\Phi \ 10^{22}$,	& $\sigma_{0.2}$& n		& $B_h$			& $T_0$,	& $T_{cl}$,	& $\sigma_{0.2}$& $n$		& $\sigma_{2c}$,& $K_I^{PV}$	& $\Phi_c^*$,& $\Phi_c^*$,	& $\sigma_2$,& $n$ 	  \\
& MPa		& MPa				& (+20),		& (+20)	& (+56),		& $^\circ$C	& $^\circ$C	& ($T_{cl}$),	& $T_{cl}$	& MPa			& ($T_{cl}$),	& m$^{-2}$	& m$^{-2}$		& ($T_S$)	& ($T_S$) \\
& 			& 					& MPa			& 		& MPa m$^{2/3}$	& 			& 			& MPa			& 			& 				& MPa$\sqrt{m}$	& 			& by (18)		& MPa		& 		  \\
\hline
\multirow{3}{*}{BM1}
&		& 0.0	& 595	& 0.060	& 		& -155	& -109	& 738	& 0.053	& 834	& 80	& 261	& 262	& 669	& 0.062 \\ \cline{3-5} \cline{7-16}
& 1400	& 7.1	& 638	& 0.058	& 24.7	& -134	& -87	& 742	& 0.052	& 843	& 81	& 303	& 308	& 720	& 0.059	\\ \cline{3-5} \cline{7-16}
&		& 15.8	& 664	& 0.057	& 		& -132	& -84	& 758	& 0.051	& 857	& 81	& 390	& 394	& 735	& 0.058	\\
\hline
\multirow{3}{*}{WM1}
& 		& 0.0	& 510	& 0.067	& 		& -120	& -93	& 714	& 0.059	& 716	& 73	& 70	& 69	& 581	& 0.068	\\ \cline{3-5} \cline{7-16}
& 1235	& 6.4	& 598	& 0.062	& 31.2	& -102	& -61	& 726	& 0.058	& 725	& 75	& 84	& 85	& 640	& 0.063	\\ \cline{3-5} \cline{7-16}
& 		& 17.5	& 593	& 0.060	& 		& -92	& -50	& 738	& 0.059	& 737	& 75	& 110	& 110	& 663	& 0.062	\\
\hline
\multirow{4}{*}{BM2}
& 		& 0.0	& 612	& 0.059	& 		& -154	& -107	& 750	& 0.052	& 848	& 81	& 520	& 499	& 688	& 0.060	\\ \cline{3-5} \cline{7-16}
& 1440	& 12.4	& 658	& 0.057	& 19.0	& -130	& -83	& 755	& 0.052	& 851	& 81	& 560	& 525	& 735	& 0.058	\\ \cline{3-5} \cline{7-16}
& 		& 19.0	& 665	& 0.056	&		& -109	& -60	& 731	& 0.053	& 827	& 80	& 352	& 325	& 742	& 0.057	\\ \cline{3-5} \cline{7-16}
& 		& 29.0	& 673	& 0.056	&		& -98	& -52	& 730	& 0.053	& 824	& 80	& 330	& 312	& 750	& 0.057	\\
\hline
\multirow{4}{*}{WM2}
&		& 0		& 545	& 0.064	& 		& -70	& -32	& 587	& 0.0613	& 676	& 72	& 24	& 30	& 617	& 0.065 \\ \cline{3-5} \cline{7-16}
& 1440	& 9.4	& 582	& 0.061	& 17.9	& -26	& +11	& 589	& 0.0612	& 678	& 72	& 35	& 34	& 657	& 0.062 \\ \cline{3-5} \cline{7-16}
& 		& 20.0	& 593	& 0.061	&		& -24	& +13	& 599	& 0.0605	& 688	& 72	& 55	& 54	& 668	& 0.062 \\ \cline{3-5} \cline{7-16}
& 		& 27.0	& 697	& 0.060	&		& -23	& +15	& 603	& 0.060		& 693	& 73	& 64	& 66	& 674	& 0.061	\\
\hline
\multirow{3}{*}{WM3}
&		& 0.0	& 483	& 0.069	& 		& -124	& -86	& 584	& 0.061	& 675	& 72	& 23	& 21	& 557	& 0.071 \\ \cline{3-5} \cline{7-16}
& 		& 12.0	& 580	& 0.619	& 41.0	& -48	& -9	& 598	& 0.060	& 686	& 72	& 30	& 28	& 650	& 0.063	\\ \cline{3-5} \cline{7-16}
& 		& 33.0	& 630	& 0.059	&		& -32	& +21	& 628	& 0.058	& 721	& 74	& 61	& 56	& 707	& 0.060 \\
\hline
\multirow{2}{*}{BM4}
& 1160	& 0.0	& 543	& 0,064	& 27.5	& -96	& -57	& 617	& 0.059	& 698	& 73	& 24	& 26	& 614	& 0.065	\\ \cline{3-5} \cline{7-16}
& 		& 8.9	& 598	& 0,060	& 		& -73	& -32	& 666	& 0.058	& 730	& 74	& 64	& 65	& 679	& 0.061	\\
\hline
\multirow{3}{*}{WM5}
& 		& 0.0	& 571	& 0.062	& 		& -97	& -56	& 633	& 0,058	& 725	& 74	& 33	& 30	& 648	& 0.063	\\ \cline{3-5} \cline{7-16}
& 1186	& 10.3	& 617	& 0.059	& 24.0	& -76	& -34	& 654	& 0.057	& 745	& 75	& 61	& 64	& 689	& 0.060 \\ \cline{3-5} \cline{7-16}
& 		& 26.0	& 644	& 0.058	&		& -59	& -16	& 666	& 0.056	& 759	& 76	& 96	& 90	& 722	& 0.059 \\
\hline
\end{tabular}
\vspace{1.5cm}\hspace{-2cm}\noindent\\ * The values of critical fluence are obtained for $\sigma_{2c}$, given in corresponding row.
\end{table}

\newpage
\renewcommand{\arraystretch}{1.5}
\begin{table}[pH] 
\caption{Mechanical properties of RPV material and characteristics of fracture of RPV wall with tolerance of $P_f=50\%$ at $J/\sigma_{0.2}=0.0365$} 
\begin{center} 
\begin{tabular}{|l|c|c|c|c|c|c|c|c|}
\hline
\multirow{3}{*}{Material}
& Fluence				& $T_{cl}$ 		& $\sigma_{0.2}$ 	& $n$ 			& $\sigma_{2c}$,	& $K_I^{PV}$			& $\Phi_c^{*}$, m$^2$	& $\Phi_c^{*}$, m$^2$ 	\\
& $\Phi \cdot 10^{22}$,	& $^{\circ}C$	& $(T_{cl})$,		& $(T_{cl})$	& MPa				& $(T_{cl})$			& 						& by (18)				\\
& m$^2$					& 				& MPa				& 				& 					& MPa$\cdot$m$^{1/2}$	& 						& 						\\
\hline
\multirow{3}{*}{BM1}
& 0.0	& -142	& 816	& 0.049	& 914	& 84	& 861	& 862	\\ \cline{2-9}
& 7.1	& -121	& 812	& 0.050	& 910	& 84	& 819	& 819	\\ \cline{2-9}
& 15.8	& -118	& 820	& 0.049	& 922	& 84	& 946	& 947	\\ \cline{2-9}
\hline
\multirow{3}{*}{WM1}
& 0.0	& -115	& 665	& 0.056	& 758	& 76	& 157	& 158	\\ \cline{2-9}
& 6.4	& -95	& 685	& 0.055	& 777	& 78	& 219	& 215	\\ \cline{2-9}
& 17.5	& -85	& 693	& 0.055	& 783	& 78	& 248	& 267	\\ \cline{2-9}
\hline
\multirow{4}{*}{BM2}
& 0.0	& -141	& 830	& 0.049	& 929	& 85	& $> 1000$	& 1794	\\ \cline{2-9}
& 12.4	& -117	& 816	& 0.049	& 914	& 85	& $> 1000$	& 1475	\\ \cline{2-9}
& 19.0	& -94	& 780	& 0.051	& 877	& 82	& 890		& 859	\\ \cline{2-9}
& 29.0	& -87	& 776	& 0.051	& 873	& 82	& 820		& 804	\\ \cline{2-9}
\hline
\multirow{4}{*}{WM2}
& 0.0	& -67	& 627	& 0.059	& 718	& 74	& 154	& 155	\\ \cline{2-9}
& 9.4	& -24	& 614	& 0.059	& 704	& 74	& 98	& 99	\\ \cline{2-9}
& 20.0	& -22	& 623	& 0.059	& 713	& 73	& 132	& 133	\\ \cline{2-9}
& 27.0	& -20	& 630	& 0.059	& 721	& 74	& 170	& 169	\\ \cline{2-9}
\hline
\multirow{3}{*}{WM3}
& 0.0	& -119	& 646	& 0.057	& 739	& 75	& 62	& 58	\\ \cline{2-9}
& 12.0	& -45	& 626	& 0.059	& 718	& 74	& 57	& 53	\\ \cline{2-9}
& 33.0	& -27	& 662	& 0.057	& 753	& 76	& 76	& 72	\\ \cline{2-9}
\hline
\multirow{2}{*}{BM4}
& 0.0	& -91	& 652	& 0.057	& 746	& 75	& 93	& 96	\\ \cline{2-9}
& 8.9	& -67	& 709	& 0.056	& 769	& 76	& 154	& 156	\\ \cline{2-9}
\hline
\multirow{3}{*}{WM5}
& 0.0	& -91	& 682	& 0.056	& 776	& 76	& 144	& 133	\\ \cline{2-9}
& 10.3	& -69	& 693	& 0.055	& 787	& 78	& 184	& 171	\\ \cline{2-9}
& 26.0	& -52	& 699	& 0.055	& 794	& 78	& 215	& 196	\\ \cline{2-9}
\hline
\end{tabular}
\end{center} 
\end{table}


\addcontentsline{toc}{section}{References}

\begin{thebibliography}{22}
\bibitem{001} PNAE G-7-002-86. Strength Calculation Norm for Nuclear Power Plant Equipment and Piping. Moscow: Energoatomizdat; 1989, 525 (in Russian).
\bibitem{002} Regulatory Guide 1.99, Revision 2. Radiation Embrittlement of Reactor Vessel Materials. U.S. Nuclear Regulatory Commission: 1988.
\bibitem{003} Unified Procedure for Lifetime Evaluation of Components and Piping in WWER NPPs "VERLIFE": 2008; 54.
\bibitem{004} AMERICAN SOCIETY OF MECHANICAL ENGINEERS, Use of Fracture Toughness Test Data to Establish Reference Temperature for Pressure Retaining Materials, Section XI, Division 1, ASME Boiler and Pressure Vessel Code Case N-629, ASME, New York: 1999; 200.
\bibitem{005} Margolin BZ, Gulenko AG, Shvetsova VA. Improved probabilistic model for fracture toughness prediction for nuclear pressure vessel steels. Int J Press Vessel Piping 1998; 75: 843–855.
\bibitem{006} Brumovsky M. Check of Master Curve application to embrittled RPVs of WWER type reactors. Int J Press Vessel Piping 2002; 79: 715–721.
\bibitem{007} Servera W, Rosinskib S, Lottc R, Kimc Ch, Weakland D. Application of Master Curve fracture toughness for reactor pressure vessel integrity assessment in the USA. Int J Press Vessel Piping 2002; 79: 701–713.
\bibitem{008} Won-Jon Yang, Bong-Sang Lee, Moo-Young Huh, Jun-Hwa Hong. Application of the local fracture stress model on the cleavage fracture of the reactor pressure vessel steels in the transition temperature region. J Nucl Mater 2003; 317: 234–242.
\bibitem{009} Kotrechko S, Mechkov Yu. A new approach to estimate irradiation embrittlement of pressure vessel steels. Int J Press Vessel Piping 2008; 85(5): 336-343.
\bibitem{010} Kotrechko S, Meshkov Yu. Limit Strength: Crystals, Metals, Structural Elements. Kiev: Naukova Dumka; 2008 (in Russian).
\bibitem{011} Kotrechko S, Meshkov Yu. Conception of Mechanical Stability of Structural Steels. Strength Mater 2009; 2: 55-78.
\bibitem{012} Beremin FM. A local criterion for cleavage fracture of a nuclear pressure vessel steel. Metall Trans 1983; 14A: 2277–2287.
\bibitem{013} Pineau A. Development of the local approach to fracture over the past 25 years: theory and applications. Int J Fract 2006; 138: 139–166.
\bibitem{014} Kotrechko S. Physical Fundamentals of Local Approach to Analysis of Cleavage Fracture. Transferability of Fracture Mechanical Characteristic. I. Dlouhy (ed), NATO Science Series. Series II. 2002; 78: 135–150.
\bibitem{015} Kotrechko S. and. Meshkov Yu. Physical fundamentals of a local approach to analysis of brittle fracture of metals and alloys. Mat Science 2001; 37(4): 583-597.
\bibitem{016} Kotrechko S. Ab-initio local approach and its use both to examine nature and to predict the irradiation effect on fracture toughness of pressure vessel steel. Pressure Vessels and Piping: Codes, Standards, Design and Analysis. Eds: Baldev Raj, B.K. Choudhary and K. Velusamy. Copyright. New Delhi, India: Narosa Publishing House; 2009. 467-477.
\bibitem{017} Kotrechko S, Strnadel B, Dlouhy I. Fracture toughness of cast ferritic steel applying local approach. Theor Appl Fract Mech 2007; 47: 171–181.
\bibitem{018} Meshkov YuYa, Pacharenko GA. Microstructure of metal and brittleness of structures. Kiev: Naukova Dumka; 1985 (in Russian).
\bibitem{019} Tanguy B, Bouchet C, Bugat S, Besson J. Local approach to fracture based prediction of the ∆T56J and ∆DT KIc; 100 shifts due to irradiation for an A508 pressure vessel steel. Eng Fract Mech 2006; 73: 191–206.
\bibitem{020} Lucon E., Walle E., Scibetta M., Chaouadi R., and Wéber M. SCK-CEN contribution to the IAEA- round robin exercise on WWER-440 RPV weld material: irradiation, annealing, and re-embrittlement. Strength Mater 2004; 36: 19-32.
\bibitem{021} Nikolaeva YuA, Nikolaev AV, Shtrombakh YaI. Radiation embrittlement of low-alloy steels. Int J Press Vessel Piping 2002; 79: 619–636.
\bibitem{022} Petti Jason P, Dodds Robert H Jr. Calibration of the Weibull stress scale parameter,  , using the Master Curve. Eng Fract Mech 2005: 72: 91–120.
\end{thebibliography}
\end{document}